\documentclass[showpacs,amsmath,notitlepage,twocolumn,superscriptaddress,nolongbibliography]{revtex4-2}
\usepackage{amsmath,amssymb,graphicx, braket, float}
\usepackage{blindtext}
\renewcommand{\thefigure}{\arabic{figure}}
\counterwithout{figure}{section}
\usepackage[export]{adjustbox}
\graphicspath{{figure/}}

\usepackage[dvipsnames]{xcolor}
\usepackage[breaklinks=true,colorlinks=true,linkcolor=red,urlcolor=blue,citecolor=blue]{hyperref}

\usepackage{graphicx, xcolor}
\usepackage[margin=.8in,text={10in,10.5in},centering]{geometry}
\usepackage{subcaption}
\usepackage[T1]{fontenc}
\usepackage{enumitem}

\usepackage[normalem]{ulem}
\newcommand{\away}[1]{\textcolor{red}{\ifmmode\text{\sout{\ensuremath{#1}}}\else\sout{#1}\fi}}

\begin{document}
	\title{Harnessing coherence generation for precision single and two-qubit quantum thermometry}

	\author{Y. Aiache} \email{youssefaiache0@gmail.com}
	\address{Laboratory of R\&D in Engineering Sciences, Faculty of Sciences and Techniques Al-Hoceima, Abdelmalek Essaadi University, BP 34. Ajdir 32003, Tetouan, Morocco}
	
	\author{A. El Allati}
	\address{Laboratory of R\&D in Engineering Sciences, Faculty of Sciences and Techniques Al-Hoceima, Abdelmalek Essaadi University, BP 34. Ajdir 32003, Tetouan, Morocco}
	\address{Max Planck Institute for the Physics of Complex Systems, Nöthnitzer Str. 38, D-01187 Dresden, Germany}
	
	\author{K. El Anouz}
	\address{Laboratory of R\&D in Engineering Sciences, Faculty of Sciences and Techniques Al-Hoceima, Abdelmalek Essaadi University, BP 34. Ajdir 32003, Tetouan, Morocco}

	%%------------------------
	\begin{abstract}
		
Quantum probes, such as single- and two-qubit probes, can accurately measure the temperature of a bosonic bath. The current investigation assesses the precision of temperature estimate using quantum Fisher information and the accompanying quantum signal-to-noise ratio. Employing an ancilla as a mediator between the probe and the bath improves thermometric sensitivity by transmitting temperature information into the probe qubit's coherences. In addition, we analyze two interacting qubits that were initially entangled or separated as quantum probes for various environmental configurations. Our findings show that increased precision is gained when the probe approaches its steady state, which is determined by the coupling between the two qubits. Furthermore, we can obtain high efficiency temperature estimation for any low temperature by changing the interaction between the two qubits.
	\end{abstract}

	\maketitle

	\section{Introduction}
\label{Intro}
In real-world circumstances, physical quantum systems interact with their surroundings, resulting in undesirable outcomes such as quantum decoherence \cite{Intro1,YA}. However, the interaction between system and reservoir can also be used to gain valuable information about quantum reservoirs. Indeed, examining and characterizing quantum reservoirs is crucial in both theoretical research and practical applications, including tasks like quantum reservoir engineering \cite{Intro2,Intro3} and coherence protection \cite{Intro4,Intro5}. Nevertheless, when dealing with a complicated quantum reservoir with many degrees of freedom, accurately estimating the reservoir's many properties becomes a significant difficulty. Quantum probes \cite{Intro8,Intro9,Intro10,Intro11,Intro12,Intro13} are an effective solution to overcome this difficulty. In fact, the use of quantum probes has lately gained popularity as a non-invasive technique for estimating parameters of interest without severely disrupting the system being studied, i.e., the quantum reservoir. The primary strategy is to place a basic quantum system, such as a qubit or a pair of qubits, in a predetermined beginning state. The system then interacts with the quantum reservoir of interest. After the interaction that imprints information about specific parameters on the state of the quantum probe, measurements are taken to extract this information \cite{Intro8,Intro14,Intro15,Intro16,Intro17}.\\

Temperature estimation is crucial for assessing complicated environments \cite{Intro21,Intro22,Intro23,Intro24}. Assessing the temperature of a quantum system is not only a fundamental task, but also of practical importance. Several quantum technologies necessitate extremely low temperatures in order to exploit sensitive non-classical properties, requiring accurate temperature estimates with little system perturbation. This requirement corresponds to the goals of quantum thermometry, representing a rich area of research that merges quantum metrology, quantum thermodynamics and open quantum systems \cite{Intro23,Intro25}. Recent research on dynamical quantum thermometry using master equations have provided insights into the non-equilibrium (equilibrium) dynamics of quantum systems \cite{RRef1,RRef2,RRef3}. Additionally, collision models have been employed to study the thermalization processes and temperature estimation in open quantum systems \cite{RRef4,RRef5,RRef6}. Driving techniques have also been explored to enhance the precision of quantum thermometry by manipulating the energy levels and transitions within the system \cite{RRef7,RRef8}.\\

The present research employs quantum probes to estimate the temperature of an environment, which is a bosonic bath, by inducing dephasing dynamics on the probe system \cite{Intro28,Intro29}. The research focuses on both single- and two-interacting qubit systems as quantum probes, optimizing their initial preparation and performing measurements to obtain temperature information. In this analysis, we apply a global Markovian master equation to investigate the dynamics of the systems. First, the single qubit probe is evaluated by comparing Quantum Fisher Information (QFI) and Quantum Signal-to-Noise Ratio (QSNR) with and without an ancillary. The probe qubit only interacts with the ancilla qubit, which dephases as a result of its contact with the sample, leaving temperature information as coherences in the probe state. Further, higher precision is achieved by increasing the probe-ancilla coupling strength. The analysis is then expanded to include two interacting qubits, taking into account both common and local baths, as well as whether the qubits are initially entangled or separated. When the system enters a steady state, the greatest accuracy of temperature estimation is achieved. Fully thermalized estimate efficiency demonstrates that optimal QSNR is determined by the coupling strength between qubits. This approach leverages the adjustable coupling to accurately control estimating efficiency at low temperatures.\\

The paper is structured as follows: In Section 2, we present the physical models for single and two-qubit systems. In Section 3, we briefly outline the tools utilized in local estimation theory and discuss our findings on the precision achieved by quantum probes in estimating the sample's temperature. The final section has a conclusion.

\section{The physical model}
\label{Sec1}
Consider a dephasing model that consists of one or two qubits interacting with a bosonic reservoir. We want to employ the qubits as quantum sensors to estimate the temperature of the thermal bath, which is assumed to be in equilibrium at temperature $ T $, using several strategies.  On the one hand, we employ a single qubit as a quantum probe, both with and without an ancillary system.  On the other hand, we investigate the usage of two interacting qubits as a quantum probe, taking into account two unique scenarios: one in which the two qubits are embedded in a common bath, and another in which each qubit interacts with its corresponding bath.

\textit{\textbf{Single qubit with ancilla-assistant}}\textthreequartersemdash We address the scenario where the probing qubit suffers from decoherence via an intermediate system, namely ancilla \textit{i.e.}, a two-level system that is directly coupled to the thermal bath. The total Hamiltonian of the probe-ancilla bath is defined as: \cite{m1}
\begin{eqnarray}
	H=H_{S}+\sum_{k}\omega_k b_k^{\dagger}b_k
	+\sigma_z\sum_{k}g_k( b_k^{\dagger}+ b_k),
\end{eqnarray} 
where $\omega_k$ is the frequency of the reservoir modes, $ b_k^{\dagger}(b_k) $ is the bosonic creation (annihilation) operator for mode $ k $ and $ g_k $ is the strength coupling of each mode with the probe qubit. The probe-ancilla Hamiltonian is given as follows
\begin{eqnarray}\label{Hpa}
	H_{S}&=&\dfrac{\hbar}{2}\omega_P\sigma^P_z+\dfrac{\hbar}{2}\omega_A\sigma^A_z\nonumber\\
	&+& \dfrac{\kappa}{2} (\sigma^{(P)}_x \sigma^{(A)}_x+\sigma^{(P)}_y\sigma^{(A)}_y).
\end{eqnarray}
The probe (ancilla) transition frequency between its ground and excited states is $ \omega_P (\omega_A) $, where $ P $ and $ A $ denote probe and ancilla, respectively. The last term defines the interaction Hamiltonian between the probe and the ancilla, with a constant coupling strength $\kappa$; $\sigma^{(j)}_x$ ($\sigma^{(j)}_y$) represents the first Pauli matrix (second Pauli matrix) of the qubit $ j=\{P,A\} $.\\

We now examine the case of two interacting qubits coupled with a thermal bath (bosonic reservoir) that functions as a quantum probe. However, two unique scenarios emerge: either the two qubits are coupled to independent local reservoirs, or they are embedded in a common reservoir.\\

\textit{\textbf{Two qubits in independent reservoirs}}\textthreequartersemdash In the case where the two-qubits are coupled to two identical and independent reservoirs with the same temperature $ T $. The total Hamiltonian is given as: 
\begin{eqnarray}
	H=H^{(1)}+H^{(2)}+H_{12}.
\end{eqnarray}
The single-qubit Hamiltonian $ H^{(i)} $, where $i = 1,2$, is expressed as
\begin{eqnarray}
	H^{(i)}&=& \dfrac{\omega^{(i)}_0}{2}\sigma^{(i)}_z+\sum_{k}\omega_k b_k^{\dagger(i)}b^{(i)}_k\nonumber\\
	&&+\sigma^{(i)}_z\sum_{k}g^{(i)}_k( b_k^{\dagger(i)}+ b^{(i)}_k),
\end{eqnarray}
$ H_{12} $ is describing the interaction between the two qubits and given by
\begin{eqnarray}\label{H12}
	H_{12}=\dfrac{\kappa}{2} (\sigma^{(1)}_x \sigma^{(2)}_x+\sigma^{(1)}_y\sigma^{(2)}_y),
\end{eqnarray}
with $\kappa$ is the coupling strength between the two qubits. Moreover, for sake of simplicity, we assume that the two qubits have the same transition frequencies $ \omega^{(i)}_0=\omega_0$.\\

\textit{\textbf{Two qubits in a common environment}}\textthreequartersemdash Next, we investigate the situation in which two interacting qubits are in contact with the same reservoir. The Hamiltonian for a bipartite qubit with a commun environment is stated as
\begin{eqnarray}\label{H_CB}
	H&=&\sum_{i=1}^{2}\dfrac{\omega^{(i)}_0}{2}\sigma^{(i)}_z+\sum_{k}\omega_k b_k^{\dagger}b_k +H_{12}\\
	&&+\sigma^{(1)}_z\sum_{k}g_{1k}( b_k^{\dagger}+ b_k)+\sigma^{(2)}_z\sum_{k}g_{2k}( b_k^{\dagger}+ b_k)\nonumber,
\end{eqnarray}
where $ H_{12} $ is defined in Eq. \ref{H12}. For simplicity, we assume that the two qubits are resonant with transition frequency $ \omega_0 $.\\

The impact of strong coupling is explored in two scenarios within the framework of a weak system-bath coupling regime: first, between the probe and the ancilla, and second, between the two qubits when viewed as a quantum probe. In fact, there are two main approaches: local and global master equations. According to references \cite{m2,m3}, the local master equations remain valid when the connection between subsystems is weak. In contrast, when the coupling is strong enough, the global master equations should be employed. To examine high coupling effects, we use a global approach. The jump operators are derived from the master equation based on the system's entire Hamiltonian eigenstates,
\begin{eqnarray}\label{ME}
	\dot{\rho}_t =-i[H_S, \rho_t]+\mathcal{L}(\rho_t),
\end{eqnarray}
where $ \mathcal{L}(\rho_t) $ is the Liouvillian and expressed as
\begin{eqnarray}
	\mathcal{L}(\rho_t)=\sum_{\omega}\gamma(\omega) \mathcal{D}[A(\omega)],
\end{eqnarray}
and $ \mathcal{D}[A(\omega)] =A(\omega) \rho_t A^{\dagger}(\omega)-\{A^{\dagger}(\omega)A(\omega), \rho_t \}/2$, with $ A(\omega) $ are the jump generators; $ \{.,.\} $ is the anti-commutator. The $ \gamma(\omega) $ are decoherence rates. A thorough derivation is provided in Appendix \ref{Append}.

\section{Thermometric performance}

Estimation theory focuses on inferring a parameter $ T $ from a set of measurement results in order to minimize estimation error. In quantum thermometry, a quantum system is utilized to estimate an unknown sample's temperature. We presume that our thermometry methodology is always separated into three major steps:
\begin{enumerate}[label=(\roman*)]
	\item preparing the probe in an appropriate state $\rho$,
	\item encoding the temperature in the probe through the probe-bath interaction, transforming the probe into a function of the temperature of the sample, denoted $\rho_T$,
	\item performing a measurement with \textbf{POVM} elements. As it is well known, the accuracy of temperature measurement is restricted to the quantum Cramér-Rao bound \cite{QCRB1,QCRB2,QCRB3}.
\end{enumerate}
\begin{eqnarray}
	\dfrac{T^2}{\delta T^2}\leq M T^2 \mathcal{F}[\rho_T]\equiv M\mathcal{R}_T.
\end{eqnarray}
Here, $ M $ is the number of measurements done, $ \delta T^2 $ is the temperature variance, and $ \mathcal{F}[\rho_T] $ is the quantum Fisher information of the quantum state $ \rho_T $ in relation to the temperature $ T $. The expression is as follows \cite{QFI,KEA,1KEA,2KEA}:
\begin{eqnarray}\label{QFI_F}
	\mathcal{F}[\rho_T]=2\sum_{k,l} \dfrac{|\braket{\phi_k|\partial_T\rho_T|\phi_l}|^2}{\lambda_k+\lambda_l},
\end{eqnarray}
$\phi_l $ and $\lambda_l $ represent the T-dependent eigenvectors and eigenvalues of $\rho_T $, respectively.We will further investigate the QSNR, $ \mathcal{R}_T = T^2 \mathcal{F}[\rho_T] $ \cite{QFI} during this research. It signifies that a greater $ R_T $ indicates better temperature sensing performance.

\subsection{Temperature sensing via quantum probes}
This section describes our findings on estimating the temperature of the sample, generally known as the thermal bath. This is done by evaluating the behavior of the QFI and QSNR for fixed values of the coupling strength between qubits, denoted as $\kappa$.

\subsubsection{Single-qubit thermometry}
For single qubit dephasing, when a quantum probe interacts directly with an Ohmic sample, the ideal initial preparation is the state 
$ \ket{+}=(\ket{0}+\ket{1})/\sqrt{2} $ \cite{Intro28}. However, for the case in which the probe qubit is interacting with the bath via the ancillary system according to their interaction Hamiltonian, we assume that the probe qubit is prepared in its excited state $ \ket{1}_P $ \cite{G_S} and we initially prepare the ancilla system in a pure state depending on the parameter $ \theta $
\begin{eqnarray}
	\ket{\psi(0)}_A =\cos(\dfrac{\theta}{2})\ket{0}_A+\sin(\dfrac{\theta}{2})\ket{1}_A. 
\end{eqnarray}
\begin{figure}[H]
	\begin{center}
		\includegraphics[scale=.8]{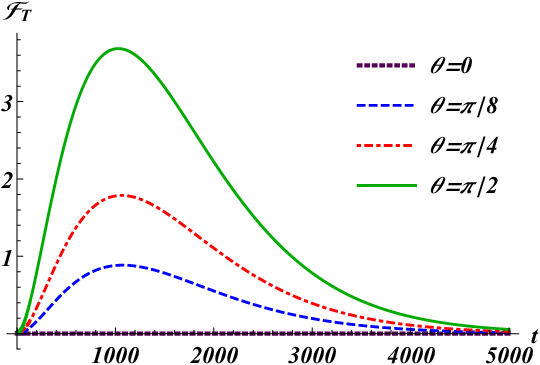}\\
		\includegraphics[scale=.8]{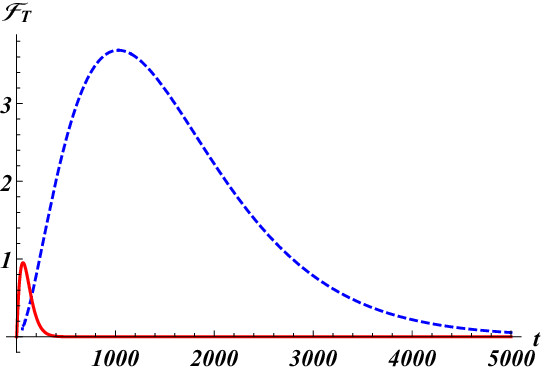}
		\caption{Top panel: QFI, $ \mathcal{F}_T$ associated with the reduced state of the probe qubit in Eq. \ref{DM_P} for different initial preparations of the ancilla for fixed $T=0.4 $, $\kappa=0.8$, $ \eta=0.01 $, and $ \Omega=10 $. Lower panel: QFI associated with probing qubit in the absence of the ancilla (red-solid-curve) and in the presence of the ancilla (blue-dashed-curve) with fixed $T=0.4$, $\kappa=0.8$, $\Omega=10$, and $\theta=\pi/2$.  All frequencies and energies are expressed in terms of $ \omega_A=\omega_P=1 $.}
		\label{fig: 3}
	\end{center}
\end{figure}

To identify the optimal initial preparation of the ancilla qubit for boosting the sensitivity of the probe qubit, we report in Fig. \ref{fig: 3} (the top panel) the QFI as a function of time interaction $ t $ for different values of the weight $\theta$. The results demonstrate that the ideal approach for the scenario where the probe qubit is indirectly connected to the bath via an ancilla system is in the excited state ($ \ket{1}_P$) and the ancilla system is in the superposition state, \textit{i.e.,} $ \ket{+}=(\ket{0}_A+\ket{1}_A)/\sqrt{2} $. However, our probe is a single qubit positioned outside the thermal bath, and we aim to do local measurements on it. As a result, the reduced state of the probe qubit is defined as
\begin{eqnarray}\label{DM_P}
	\rho_p(t)=Tr_A\big\{ \rho_{PA}(t) \big\}=\begin{pmatrix}
		(1+\mathcal{W})/2 & \mathcal{X} \\
		\mathcal{X}^* & (1-\mathcal{W})/2
	\end{pmatrix}.\nonumber\\
\end{eqnarray} 
where the population (diagonal) and coherences (off-diagonal) terms are specified respectively by
\begin{eqnarray}\label{E_P}
	\mathcal{W}&=&\frac{1}{4} \left(-2+\left(1+e^{4 i \kappa t}\right) (\sinh (\mathcal{B}~t)-\cosh (\mathcal{B}~t))\right)\nonumber\\
	\mathcal{X}&=&\frac{1}{4} \left(e^{-\mathcal{Z}_2~t}-e^{i \mathcal{Z}_1~t}\right),
\end{eqnarray}
where
\begin{eqnarray}
	\mathcal{B}&=&2 \kappa  \left(\pi  \eta  e^{-\frac{2 \kappa}{\Omega }} \coth \left(\frac{\kappa}{T}\right)+i\right),\nonumber\\
	\mathcal{Z}_1&=& i \pi  \eta  \kappa e^{-\frac{2 \kappa}{\Omega }} \left(\coth \left(\frac{\kappa}{T}\right)-1\right)+\kappa-1,\nonumber\\
	\mathcal{Z}_2&=& \pi  \eta  \kappa e^{\kappa \left(\frac{1}{T}-\frac{2}{\Omega }\right)} \text{csch}\left(\frac{\kappa}{T}\right)+i (\kappa+1).
\end{eqnarray}
\begin{figure}[H]
	\begin{center}
		\includegraphics[scale=.73]{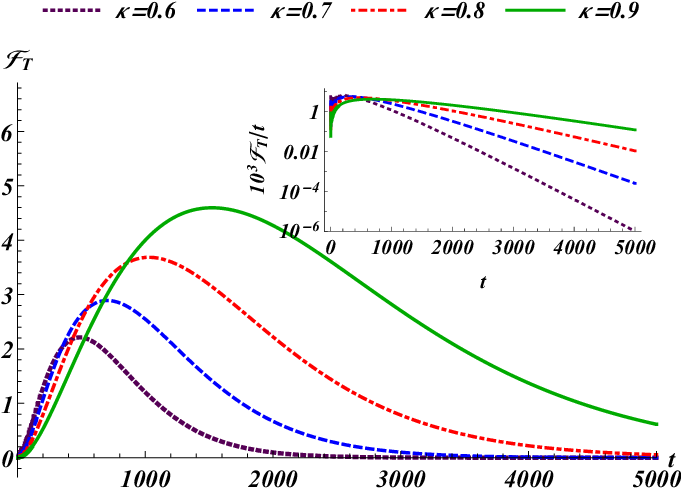}\\
		\includegraphics[scale=.454]{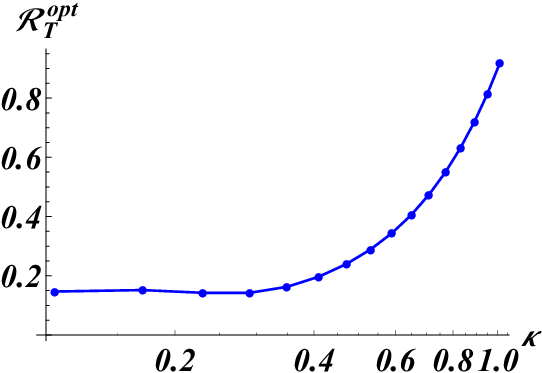}
		\includegraphics[scale=.454]{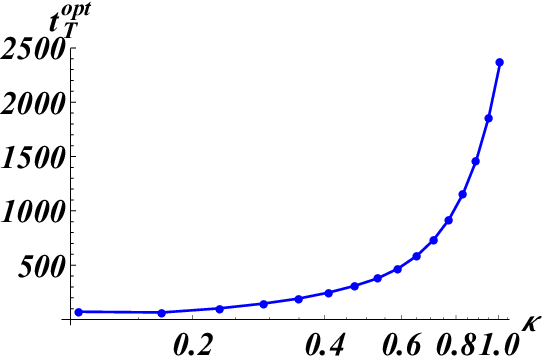}
		\caption{Top panel: QFI and QFI/t (inset) of the state specified in Eq. \ref{DM_P} vs time interaction $ t $ for different probe-ancilla coupling strengths $ \kappa $ . Lower panel: The left plot shows the highest value of QSNR vs $\kappa$. We plot the optimal time, \textit{i.e.}, the moment when QSNR reaches the highest value, as a function of $ \kappa $ to ensure proper behavior. We set $ T = 0.4 $, $ \eta = 0.01$, and $ \Omega = 10$. All frequencies and energies are expressed in terms of $ \omega_A=\omega_P=1 $.}
		\label{fig:4}
	\end{center}
\end{figure}
The coherences and population terms of the probe qubit include critical information about the sample's temperature. This indirect coupling via the ancilla alters the quantum state of the probe qubit, causing its coherences and population dynamics to reflect the sample's thermal parameters. Thus, seeing these quantum features offers information on the sample's temperature within this framework.

In Figure \ref{fig: 3} (the lower panel), we compare the QFI associated with the probe being directly linked to the bath or uncoupled from the bath and interacting with an ancilla. Furthermore, it is believed that the ancilla is connected directly to the sample. The findings are plotted as a function of interaction time $t$. At first, the probe-bath situation, where QFI achieves its maximum information, holds the majority of the thermometric information; but, as time passes, the information in the indirect coupling scenario increases. Thus, both situations provide equivalent information, and over time, the indirect coupling with considerably higher sensitivity holds the majority of the information about the sample's temperature.\\

Obviously, the ability of the probe qubit in the presence of an intermediate system (ancilla) to accumulate information for a very long time allows it to reach a significantly higher thermometric sensitivity. An interesting feature of this strategy is that, after a certain initial time, a local measurement on the probe qubit is able to extract substantially all the information from the state. This allows the thermometer to be implemented more easily. Thus, for the rest of the manuscript, we shall focus on the behavior of QFI and QSNR, associated with the reduced state defined in Eq. \ref{DM_P}.

\begin{figure}[H]
	\begin{center}
		\includegraphics[scale=.8]{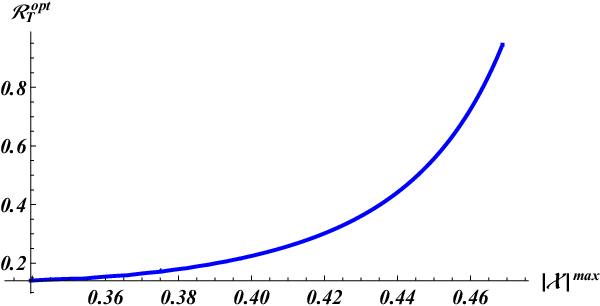}			
		\caption{Parametric plot of optimal QSNR versus maximal coherence generated in the probe qubit \cite{Youssef2}. We set $\eta=0.1 $, $T=0.4 $ and $ \Omega=10 $.  All frequencies and energies are in units of $ \omega_A=\omega_P=1 $.}
		\label{fig:7}
	\end{center}
\end{figure}

Fig. \ref{fig:4} illustrates how the coupling strength between our probe and the ancilla affects the sensitivity of temperature estimate using QFI, $ \mathcal{F}_T$. We can clearly see that increasing the constant coupling probe-ancilla improves the sensitivity of temperature estimation. Besides, we plot QFI per unit time ($\mathcal{F}_T/t$) versus time t \cite{QFIt} (shown in the inset of Fig. \ref{fig:4}), for various coupling strengths ($\kappa$). In short-time interaction, the sensitivity of $\mathcal{F}_T/t$ is robust, but decreases sharply over time. Particularly, stronger coupling maintains robust values of $\mathcal{F}_T/t$. Indeed, $\mathcal{F}_T/t$ for  $\kappa=0.9$ (green line) is notably more persistent than $\kappa=0.6$. This indicates that stronger coupling could optimize precision in quantum estimation and measurement over time. Furthermore, in the lower panel of Fig.  \ref{fig:4} (left plot), we report the highest value of the QSNR, referred to as $\mathcal{R}_T^{opt}$, as a function of the coupling strength $\kappa$. As the coupling strength $\kappa$ grows, so does the value of $\mathcal{R}_T^{opt}$. Our findings indicate that a strong coupling between the probe and ancilla improves thermal sensitivity and accuracy in temperature estimation.\\

The time corresponding to the maximum of QSNR, $\mathcal{R}_T^{opt}$, is called optimal time, $t^{opt}$. Fig. \ref{fig:4} shows that higher coupling $\kappa$ values lead to longer optimum times. We discover a trade-off between optimal QSNR and optimal encoding time, determined by the coupling strength $\kappa$. A higher optimal QSNR necessitates a longer optimal encoding duration.\\

Fig. \ref{fig:7} depicts the relationship between optimal QSNR and maximum generated coherences ($ \mathcal{X}^{max} $) in the probing qubit. It is evident that increasing the maximum of created coherences significantly increases the probe's thermal sensitivity. As a result, the ancilla converts fundamentally coherent information about the bath temperature into coherences in the probe qubit. In fact, Reference \cite{Asghar} found that adding an ancilla as an intermediate increased the probe's thermal sensitivity and improved steady-state estimation efficiency.

\begin{figure}[H]
	\begin{center}
		\includegraphics[scale=.8]{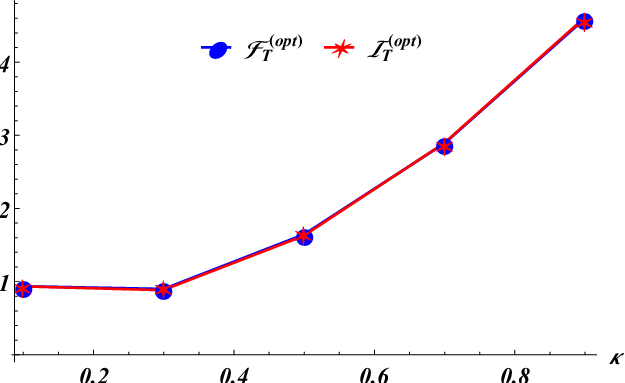}			
		\caption{ The optimal Fisher information $ \mathcal{I}_T $ and optimal quantum Fisher information $ \mathcal{F}_T $ depend on the probe-ancilla coupling. We used $\eta=0.01 $, $T=0.4 $, and $ \Omega=10 $.  All frequencies and energies are expressed in terms of $ \omega_A=\omega_P=1 $.}
		\label{fig:QFI_vs_CFI}
	\end{center}
\end{figure}

Determining the best measurement is critical for getting the maximum possible precision of the estimated parameter in practical experiments. However, for a single qubit, the Fisher information related to the measurement can be computed as \cite{QFIt,Y4}
\begin{eqnarray}
	\mathcal{I}_T=\dfrac{1}{\braket{\Delta X^2}}\bigg(\dfrac{\partial\braket{X}}{\partial T}\bigg)^2,
\end{eqnarray}
where $ \braket{X} $ and $ \braket{\Delta X^2} $ represent the mean and variance of the measured observable $ X $. The QFI represents the upper limit of the Fisher information associated with the measurement $ X $, i.e., 
$$ \mathcal{F}_T=\max_X \mathcal{I}_T=Tr [\rho_p(t).\Lambda.\Lambda],$$ 
where $ \Lambda $ is the symmetric logarithmic derivative (SLD). For any mixed state, the SLD can be written as \cite{SLD}
\begin{eqnarray}
	\Lambda=2\partial_T \rho_p(t)-\dfrac{\partial_T \mathcal{P}}{1-\mathcal{P}}\sigma_y\rho_p^t(t)\sigma_y,
\end{eqnarray}
with $ \mathcal{P} $ is the purity defined as $ \mathcal{P}=Tr[\rho_p(t)^2] $, and $ \rho_p^t(t) $ is the transposed density matrix $ \rho_p(t) $. Since our probe is a single qubit, hence, we can re-write the density matrix of the probe qubit (Eq. \ref{DM_P}) in the Bloch representation \textit{i.e.,} $\rho_p(t)=(\mathbb{I}_2+\mathbf{r} \cdot \hat{\boldsymbol{\sigma}})/2$, where $ \mathbb{I}_2 $ denotes the $2 \times 2$ identity matrix, $\mathbf{r} = (r_x, r_y, r_z)$ is the real Bloch vector, and $\hat{\boldsymbol{\sigma}} = (\hat{\sigma}_x, \hat{\sigma}_y, \hat{\sigma}_z)$ denotes the Pauli matrices. Using Eq. \ref{DM_P}, we can easily obtain the elements of Bloch’s vector as follows:
\begin{equation}\label{Bloch}
	r_x=\mathcal{X}+\mathcal{X}^*, \quad r_y=i (\mathcal{X}-\mathcal{X}^*), \quad r_z=\mathcal{W},
\end{equation}
where $ \mathcal{X} $ and $ \mathcal{W} $ are defined in Eq. \ref{E_P}. Using the above equation, we can straightforwardly derive the SLD as
\begin{eqnarray}\label{SLD}
	\Lambda=c_0 \mathbb{I}_2+c_x\sigma_x+c_y\sigma_y+c_z\sigma_z,
\end{eqnarray}
where
\begin{eqnarray}
	c_0&=&\frac{\partial_T \mathcal{P}}{2 (\mathcal{P}-1)},\;\;\;\;\;\;\;\;\;\;\;\; c_x=\frac{r_x \partial_T\mathcal{P}}{2-2 \mathcal{P}}+\partial_Tr_x,\nonumber\\
	c_y&=&\frac{r_y \partial_T\mathcal{P}}{2-2 \mathcal{P}}+\partial_Tr_y,\;\;\; c_z=\frac{r_z \partial_T\mathcal{P}}{2-2 \mathcal{P}}+\partial_Tr_z.
\end{eqnarray}
The coefficients of the decomposition vary with temperature, time, and coupling strength, while the projectors are independent of all these parameters. However, by calculating the QFI using the formula \ref{QFI_F} and using the SLD (Eq. \ref{SLD}), one can easily obtain
\begin{eqnarray}
	\mathcal{F}_T&=&Tr [\rho_p(t).\Lambda.\Lambda]\nonumber\\
	&=&\frac{(\partial_T\mathcal{P})^2 \left(1-|r|^2\right)}{4 (\mathcal{P}-1)^2}+(\partial_T r_x)^2+(\partial_Tr_y)^2+(\partial_Tr_z)^2,\nonumber\\
\end{eqnarray} 
where $ |r|^2=r_x^2+r_y^2+r_z^2 $ and $ \mathcal{P}=(1+|r|^2)/2 $.\\

By calculating the symmetric logarithmic derivative, we determine the optimal measurement to be applied to the probe qubit for inferring the temperature in a bosonic environment. However, in the context of probe-ancilla coupling, we illustrate in Fig. \ref{fig:QFI_vs_CFI} the optimal QFI and the optimal Fisher information associated with the measurement  using $ \sigma_x $, \textit{i.e.}, $ \mathcal{I}_T = (\partial_T r_x)^2/(1 - r_x^2) $. Clearly, we observe that the maximum Fisher information coincides with the optimal QFI. This demonstrates that excellent temperature estimates at the quantum limit can be achieved using a practical method that incorporates specialized measurement techniques after interaction with the thermal sample.

\subsubsection{Two-qubit thermometry}

Let's compare the performance of a two-qubit quantum probe for temperature estimation while always paying attention to QFI and QSNR behavior. We study the influence of qubit-qubit interaction as well as probe interaction with a common (independent) bath. The concepts of sharing the same bath or having two independent duplicates of the same bath do not require different physical systems. In a practical setting, these situations correspond to the qubits being near together and interacting with the same piece of the environment, or being far away with negligible spatial correlations of the bath. We take the following initial state for the two interacting qubits:
\begin{eqnarray}\label{TQS}
	\ket{\psi(0)}=\cos(\dfrac{\theta}{2})\ket{01}+\sin(\dfrac{\theta}{2})\ket{10}.
\end{eqnarray}

\begin{figure}[H]
	\begin{center}
		\includegraphics[scale=.9]{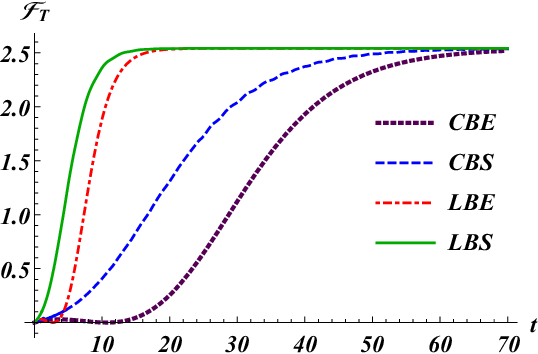}
		\caption{QFI for independent and common baths against $t$ using four different initial sittings of the qubits: two qubits initially in an entangled or separable state in a common bath, and two qubits in independent baths initialized in an entangled or separable state. For fixed $ T =0.4$, $\kappa=0.6$, $ \eta_1=0.01 $ and $ \eta_2=0.05 $. All the frequencies and the energies are in units of $ \omega_1=\omega_2=1 $.}
		\label{fig: T_Q_D}
	\end{center}
\end{figure}

The plots in Fig. \ref{fig: T_Q_D} exhibit the fluctuation of the nonequilibrium QFI, $ \mathcal{F}_T $, with interaction time $ t $. We observe that nonequilibrium conditions don't improve the precision of temperature estimate for the two-qubit system utilized as a quantum probe. The peak value of $ \mathcal{F}_T $ occurs as the thermometer enters a steady state. Figure \ref{fig: T_Q_D} indicates that a thermometer with two discrete qubits in contact with independent thermal baths reaches the maximum QFI faster than other configurations.  Additionally, entanglement does not appear to greatly improve the probe's sensitivity.\\

Given that the maximum QFI for two interacting qubits is obtained under equilibrium conditions with the same steady value in both common and local bath situations, regardless of whether the qubits are initially entangled or separated, we need to assess the probe's performance. We will accomplish this by measuring the QSNR over long encoding durations. This investigation seeks to provide further information about the probe's efficiency for temperature estimation. Thus, the steady state of the two qubits is as follows:
\begin{eqnarray}\label{DM_Steady}
	\rho(\infty)&=&\dfrac{1}{2}\big(\ket{01}\bra{01}+\ket{10}\bra{10}\big)\nonumber\\
	&=&\frac{1}{2} \tanh \left(\frac{-k}{T}\right)\big(\ket{01}\bra{10}+\ket{10}\bra{01}\big).
\end{eqnarray}
According to the above expression, the steady state QFI can be obtained as 
\begin{eqnarray}\label{F_S}
	\mathcal{F}_T(t\longrightarrow\infty)&=&\mathcal{F}_T(\infty)= \frac{2 \kappa ^2}{T^4 \left(\cosh \left(\frac{2 \kappa }{T}\right)+1\right)}.
\end{eqnarray}
To infer the value of temperature in a bosonic environment in possible experimental implementations, we use the POVM on the two qubits. By improving the measurement method, we may make our quantum thermometry technique more practical, ensuring that the experimental procedures are practicable and robust under realistic conditions. In this context, the Fisher information is provided as 
\begin{eqnarray}
	\hbox{F}^C(T)=\sum_{i=1}^{4}\dfrac{\big[\partial_T p_i\big]^2}{p_i},
\end{eqnarray}
$ p_i $ represents the probability distribution of the alternative outcomes. Hence, for density matrix Eq. \ref{DM_Steady}, it is straightforward to obtain 
$$ \hbox{F}^C(T)=\frac{2 \kappa ^2}{T^4 \left(\cosh \left(\frac{2 \kappa }{T}\right)+1\right)} ,$$
 which is precisely equivalent to the QFI provided in Eq. \ref{F_S}.\

\begin{figure}[H]
	\begin{center}
		\includegraphics[scale=.9]{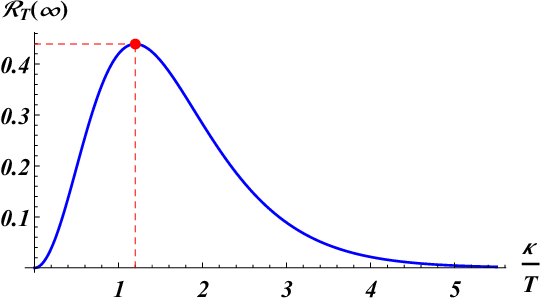}\\
		\includegraphics[scale=.9]{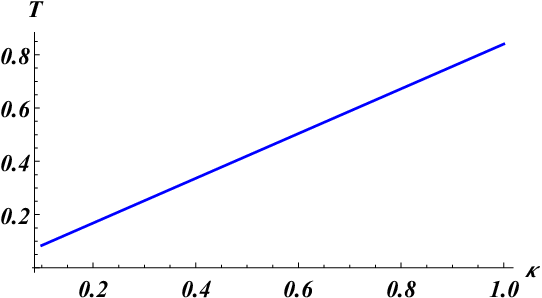}
		\caption{The top panel shows how the steady-state QSNR, indicated as $\mathcal{R}_T(\infty)$, relates to the ratio $\kappa/T$. The optimal QSNR value is highlighted by red-point. The lower panel shows the link between $\kappa$ and $T$ at this ideal position. All frequencies and energies are expressed in terms of $ \omega_1=\omega_2=1 $.}
		\label{fig:OPtim}
	\end{center}
\end{figure}

The optimal QSNR ($ \mathcal{R}_T(\infty) $) in lengthy encoding times can be obtained when $  \tanh \left(\frac{\kappa }{T}\right)=\frac{T}{\kappa} $, which is around $0.44$. However, Fig. \ref{fig:OPtim} (top panel) shows the relationship between steady-state QSNR and the ratio of qubit-qubit coupling ($\kappa$) to temperature ($ T $). A significant insight emerges: the steady-state $ \mathcal{R}_T(\infty) $ is solely defined by the ratio of $ \kappa/T $, regardless of the intensity of the dipole interactions. In other words, the optimal steady QSNR depends simply on the relative values of coupling $ \kappa $ and temperature $ T $, not their individual magnitudes. This implies that by adjusting the coupling strength between the qubits to approximately $ \kappa/T \approx 1.2 $, the sensitivity of temperature sensing can be enhanced. This enhancement is shown in lower pane in Fig. \ref{fig:OPtim}. Optimizing the ratio of qubit-qubit coupling to temperature increases the quantum thermometer's temperature sensing capacity, particularly at low temperatures. The aim is to strike the correct balance between coupling strength and temperature to enhance QSNR and temperature measurement sensitivity.

\section{Conclusions}
In this paper, we investigated the temperature estimation of a bosonic reservoir utilizing single- and two-qubit quantum probes. The thermal bath is viewed as a collection of non-interacting bosonic modes. We examined the QFI and QSNR of the probes in various beginning states.\\

On the one hand, we compared the thermometric performance of a single qubit with and without an ancilla system, which acts as a bridge between the probe qubit and the sample under consideration. Specifically, the ancilla is coupled to the probe qubit via an interaction Hamiltonian and is directly coupled to the bath. Our findings show that, over short periods of time, a qubit directly connected to the bath is the most effective approach to probe its temperature. However, the existence of the ancilla system results in far higher precision, as the ancilla imprints information from the sample and subsequently maps it into coherences in the probe system. 
Furthermore, by adjusting the probe-ancilla coupling strength, the ancilla system enabled the probe state to be significantly more temperature sensitive than the probe state alone.\\

In this study of two qubits, two situations were considered: one with each qubit interacting with its own independent bath, and another with both qubits interacting with a common bath. The performance of entangled and separated qubits was compared. The findings revealed that nonequilibrium situations do not increase temperature estimation precision, since the QFI for all techniques converges to the same steady value. Initial correlations between qubits do not improve probe performance, and maximal accuracy is obtained in a relatively short period when two qubits in independent local baths are separable.  Further research into steady-state efficiency demonstrated that varying the qubit-qubit interaction intensity enables high-efficiency temperature sensing at low temperatures. The proposed approach has substantial applications in high-resolution quantum thermometry.\\

For potential realizations, superconducting qubits \cite{Sup1,Sup2}, quantum dots \cite{QD1,QD2}, and Bose-Einstein condensates (BECs) \cite{BEC1,BEC2} are promising platforms. In superconducting qubits, transmons or flux qubits could serve as the probe, with other qubits or resonators as the ancilla, and resistive elements as the thermal bath. Quantum dots and BECs can similarly offer flexible and precise environments for implementing our quantum thermometer.

\renewcommand{\theequation}{A-\arabic{equation}}
\renewcommand{\thefigure}{A-\arabic{figure}}
% redefine the command that creates the equation no.
\appendix
\setcounter{equation}{0}  % reset counter
\setcounter{figure}{0}    
\section{Derivation of the global master equation}
\label{Append}

The global master equation in Eq. \ref{ME} is derived using the methods described in Refs. \cite{Intro1,m2} for open quantum systems. To begin, we define the eigenvalues and eigenvectors of $ H_S $, or the total Hamiltonian of the two qubits, as follows:
\begin{eqnarray}
	H_S\ket{E_i}=E_i\ket{E_i},
\end{eqnarray}
such that the associated projection operators are
\begin{eqnarray}
	\Pi(i)=\ket{E_i}\bra{E_i}.
\end{eqnarray}
The interaction Hamiltonian between the system-bath can be written in a general form
\begin{eqnarray}
	H_I=A\otimes B,
\end{eqnarray}
where $A$ and $B$ are respectively,  the system’s and bath’s operators. In particular, for the bosonic bath we have $ B= \sum_{k}g_k( b_k^{\dagger}+ b_k)$. The jump operators of the system are defined as 
\begin{eqnarray}\label{Jump}
	A(\omega)=\sum_{E_m-E_n=\omega} \Pi(n)\;A\;\Pi(m),
\end{eqnarray}
and obeys the following relations
\begin{eqnarray}
	\big[A(\omega),H_S\big]&=&\omega A(\omega),\nonumber\\
	\big[A^{\dagger}(\omega),H_S\big]&=&-\omega A^{\dagger}(\omega).
\end{eqnarray}
The sum in Eq. \ref{Jump} includes all energy eigenvalues $ E_m $ and $ E_n $ of $ H_S $ with a constant energy difference of $ \omega = E_m - E_n $. In Eq. \ref{ME}, the decoherence rates are calculated by following Refs \cite{Intro1,Rivas} as follows
\begin{eqnarray}
	\gamma(\omega)=2 \;Re\biggl[\int_{0}^{\infty}d\tau\; e^{i\omega\tau} \braket{e^{iH_B\tau} B e^{-iH_B\tau} B}\biggr].\nonumber\\
\end{eqnarray}
By substituting the form of $ B $, we straightforwardly obtain 
\begin{eqnarray}
	\gamma(\omega)=&2 \;Re\biggl[\int_{0}^{\infty}d\tau\; e^{i\omega\tau}\sum_{k}|g_k|^2 (\braket{b_kb_k^{\dagger}}e^{-i\omega_k\tau}\nonumber\\
	&+\braket{b^{\dagger}_k b_k}e^{i\omega_k\tau})\biggr].
\end{eqnarray}
Introducing the spectral density $ J(\omega) =\sum_{k}|g_k|^2 \;\delta(\omega-\omega_k)$, and assuming that the bath is in a thermal state with a temperature $ T $, we can write the global master equation as
\begin{eqnarray}\label{GME}
	\dot{\rho}_t =&-&i[H_S, \rho_t]\nonumber\\
	&+&\sum_{\omega}2\pi J(\omega)n(\omega)\big[A^{\dagger}(\omega) \rho_t A(\omega)\nonumber\\
	&-&\{A(\omega)A^{\dagger}(\omega), \rho_t \}/2\big]\nonumber\\
	&+&\sum_{\omega}2\pi J(\omega)\big[n(\omega)+1\big]\big[A(\omega) \rho_t A^{\dagger}(\omega)\nonumber\\
	&-&\{A^{\dagger}(\omega)A(\omega), \rho_t \}/2\big],
\end{eqnarray}
where $ n(\omega) $ is the average thermal excitation number and expressed as follows
\begin{eqnarray}
	n(\omega)=\big(e^{\omega/T}-1\big)^{-1}.
\end{eqnarray}
We will investigate the thermometry of the bath spectral density, with a particular emphasis on the Ohmic form with a cutoff frequency $ \Omega $, which is given as \cite{Intro1}
\begin{eqnarray}
	J(\omega)=\eta\omega e^{-\omega/\Omega}.
\end{eqnarray}
To  derive the case where the probe qubit indirectly coupled to the bath via an ancilla system that is directly coupled to the bath, we simply replace $ A (\omega)$ by $ \sum_{E_m-E_n=\omega} \Pi(n)\;\sigma_z^A\;\Pi(m) $ in Eq. \ref{GME}, where $\sigma_z^A =\mathbb{I}\otimes\sigma_z $.\\

In the case of two coupled qubits acting as a quantum probe, we distinct two scenarios \textit{i.e.,} the two interacting qubits are coupled to a common or each qubit is coupled to its local bath. However, we begin with the case where each qubit is coupled to its local bath. Since we are dealing with master equations in the Born-Markov regime, the dissipators are additive \cite{Diss}. Therefore, the global master equation of the two interacting qubits coupled to the local bath can be expressed as follows: 
\begin{eqnarray}\label{GME-LB}
	\dot{\rho}_t =-i[H_S, \rho_t]+\mathcal{L}_1(\rho_t)+\mathcal{L}_2(\rho_t),
\end{eqnarray}
where $ \mathcal{L}_j(\rho_t)$ with $ j=\{1,\;2\} $, describe the dissipation of the qubit $ j $ due to its local thermal bath, and have the following form
\begin{eqnarray}
	\mathcal{L}_j(\rho_t) &=&\sum_{\omega}2\pi J_j(\omega)n_j(\omega)\big[A_j^{\dagger}(\omega) \rho_t A_j(\omega)\nonumber\\
	&-&\{A_j(\omega)A_j^{\dagger}(\omega), \rho_t \}/2\big]\nonumber\\
	&+&\sum_{\omega}2\pi J_j(\omega)\big[n_j(\omega)+1\big]\big[A_j(\omega) \rho_t A_j^{\dagger}(\omega)\nonumber\\
	&-&\{A_j^{\dagger}(\omega)A_j(\omega), \rho_t \}/2\big],
\end{eqnarray}
where $ 	J_j(\omega)=\eta_j\;\omega\; e^{-\omega/\Omega}$ with $ \eta_j $ is the constant coupling between the qubit $ j $ and its associated bath; $ 	n_j(\omega)=\big(e^{\omega/T_j}-1\big)^{-1} $ is the thermal occupation number of bath $ j $. The jump operators of the first and second qubit are $ A_1 (\omega)= \sum_{E_m-E_n=\omega} \Pi(n)\;(\sigma_z\otimes\mathbb{I})\;\Pi(m) $ and $A_2 (\omega)= \sum_{E_m-E_n=\omega} \Pi(n)\;(\mathbb{I}\otimes\sigma_z)\;\Pi(m)$, respectively.\\

In contrast to local baths, the common reservoir shared by the two qubits will introduce dissipative terms, as is evident from Eq. \ref{H_CB}. These terms include those of the form given in Eq. \ref{GME-LB}, as well as cross terms. Hence,
\begin{eqnarray}\label{GME-CB}
	\dot{\rho}_t =-i[H_S, \rho_t] +\mathcal{L}^C_1(\rho_t)+\mathcal{L}^C_2(\rho_t)+\mathcal{L}^C_{12}(\rho_t),
\end{eqnarray}
here the index $ C $ reflecting the common bath.  The dissipators arising from the common baths are given as
\begin{eqnarray}
	\mathcal{L}^C_j(\rho_t) &=&\sum_{\omega}2\pi J_j(\omega)n_C(\omega)\big[A_j^{\dagger}(\omega) \rho_t A_j(\omega)\nonumber\\
	&-&\{A_j(\omega)A_j^{\dagger}(\omega), \rho_t \}/2\big]\nonumber\\
	&+&\sum_{\omega}2\pi J_j(\omega)\big[n_C(\omega)+1\big]\big[A_j(\omega) \rho_t A_j^{\dagger}(\omega)\nonumber\\
	&-&\{A_j^{\dagger}(\omega)A_j(\omega), \rho_t \}/2\big],
\end{eqnarray}
 and $ \mathcal{L}^C_{12}(\rho_t) $ reflects the collective behavior of the two qubits induced by the common bath, and expressed as
\begin{eqnarray}
 	\mathcal{L}^C_{12}(\rho_t) &=&\sum_{\omega}2\pi\; \Gamma^C_{12}(\omega)\bigg(n_C(\omega)\big[A_1^{\dagger}(\omega) \rho_t A_2(\omega)\nonumber\\
 	&-&\{A_2(\omega)A_1^{\dagger}(\omega), \rho_t \}/2\big]\nonumber\\
 	&+&\big[n_C(\omega)+1\big]\big[A_2(\omega) \rho_t A_1^{\dagger}(\omega)\nonumber\\
 	&-&\{A_1^{\dagger}(\omega)A_2(\omega), \rho_t \}/2\big]\nonumber\\
 	&+&n_C(\omega)\big[A_2^{\dagger}(\omega) \rho_t A_1(\omega)\nonumber\\
 	&-&\{A_1(\omega)A_2^{\dagger}(\omega), \rho_t \}/2\big]\nonumber\\
 	&+&\big[n_C(\omega)+1\big]\big[A_1(\omega) \rho_t A_2^{\dagger}(\omega)\nonumber\\
 	&-&\{A_2^{\dagger}(\omega)A_1(\omega), \rho_t \}/2\big]\bigg),
\end{eqnarray}
with $ \Gamma^C_{12}(\omega)=\sqrt{J_1(\omega)\;J_2(\omega)} $. The thermal occupation number of common bath $ n_C(\omega) =\big(e^{\omega/T_C}-1\big)^{-1} $ with $ T_C $ is the temperature of the common bath.
 
\section*{Acknowledgment}

A.E.A. would like to thank the Max Planck Institute for the Physics of Complex Systems for the financial support and the friendly environment. The authors thank Prof. Bassano Vacchini for useful discussions and careful reading of the manuscript.

The authors declare that they have no conflict of interest.

\end{document}